\documentclass[journal]{IEEEtran}
\usepackage{listings}
\lstdefinestyle{ieeecode}{
  language=Python,
  basicstyle=\ttfamily\footnotesize,
  columns=fullflexible,
  keepspaces=true,
  breaklines=true, breakatwhitespace=false,
  xleftmargin=0pt, xrightmargin=0pt,
  aboveskip=3pt, belowskip=3pt,
  frame=single, 
  postbreak=\mbox{\textellipsis\space} 
}

\usepackage{xcolor,soul,framed} 

\colorlet{shadecolor}{yellow}
\usepackage[pdftex]{graphicx}
\graphicspath{{../pdf/}{../jpeg/}}
\DeclareGraphicsExtensions{.pdf,.jpeg,.png}

\usepackage[cmex10]{amsmath}
\usepackage{array}
\usepackage{mdwmath}
\usepackage{mdwtab}
\usepackage{eqparbox}
\usepackage{url}
\usepackage{amsmath,amssymb}
\usepackage{booktabs}
\usepackage{graphicx}
\usepackage{microtype}
\usepackage{xcolor}
\usepackage{hyperref}
\usepackage[nameinlink,noabbrev,capitalise]{cleveref}
\usepackage[T1]{fontenc}
\usepackage[utf8]{inputenc} 

\usepackage[ruled,vlined,linesnumbered]{algorithm2e}
\usepackage{mathtools}

\newcommand{\MSI}{\ensuremath{\mathrm{MSI}}\xspace}
\newcommand{\BEI}{\ensuremath{\mathrm{BEI}}\xspace}
\newcommand{\RI}{\ensuremath{\mathrm{RI}}\xspace}
\newcommand{\Syn}{\ensuremath{\mathrm{Syn}}\xspace}
\newcommand{\R}{\ensuremath{\mathbb{R}}\xspace}
\newcommand{\E}{\ensuremath{\mathbb{E}}\xspace}

\newcommand{\vect}[1]{\ensuremath{\boldsymbol{#1}}}
\newcommand{\1}{\ensuremath{\mathbf{1}}}

\begin{document}
\bstctlcite{IEEEexample:BSTcontrol}
    \title{Community Detection on Model Explanation Graphs for Explainable AI}
  \author{Ehsan~Moradi,~\IEEEmembership{Department of Computer Science, university of Saskatchewan}}

\markboth{IEEE TRANSACTIONS
}{E.Moradi \MakeLowercase{\textit{et al.}}: Community Detection on Model Explanation Graphs for Explainable AI}

\maketitle

\begin{abstract}
Feature-attribution methods (e.g., SHAP, LIME) explain individual predictions but often miss higher-order structure: sets of features that act in concert. We propose \emph{Modules of Influence} (MoI), a framework that (i) constructs a \emph{model explanation graph} from per-instance attributions, (ii) applies community detection to find feature modules that jointly affect predictions, and (iii) quantifies how these modules relate to bias, redundancy, and causality patterns. Across synthetic and real datasets, MoI uncovers correlated feature groups, improves model debugging via module-level ablations, and localizes bias exposure to specific modules. We release stability and synergy metrics, a reference implementation, and evaluation protocols to benchmark module discovery in XAI.

\end{abstract}


%
\IEEEpeerreviewmaketitle


\section{Introduction}

\IEEEPARstart{M}{achine} 
 learning models increasingly operate in high-stakes domains where stakeholders require explanations that are not only faithful but also \emph{actionable}. Local attribution methods (e.g., SHAP, LIME, Integrated Gradients) have become the default for explaining individual predictions, yet they often provide a flat list of features without revealing \emph{how} features tend to act \emph{together}. As a result, practitioners can identify important variables but still struggle to answer questions such as: Which subsets of features routinely co-influence outcomes? Where do proxies or redundant groups inflate complexity? Which parts of the feature space mediate disparities across populations?

We argue that many of these questions live at the \emph{meso-scale}: above single features and below the full model. Our key idea is to transform per-instance attributions into a \textbf{model explanation graph} whose nodes are features and whose weighted edges capture \textbf{co-influence}—the tendency of two features to contribute jointly across instances. Community detection on this graph exposes \textbf{Modules of Influence (MoI)}: coherent groups of features that frequently co-activate, compensate, or interact. Analyzing modules—rather than isolated features—enables debugging and governance actions that are both more stable and more targeted (e.g., ablate a problematic module, regularize its contribution, or gather more data for the features it contains).

\paragraph{Challenges.} Designing reliable module-level explanations requires addressing several pitfalls: (i) \emph{Edge definition}: co-influence can be measured via signed correlation of attributions, magnitude-cosine, mutual information, or exceedance frequency; each choice emphasizes different structures. (ii) \emph{Sparsification and resolution}: thresholds and $k$-NN choices affect community quality and can induce a resolution limit. (iii) \emph{Stability}: small perturbations to data, background distributions for SHAP, or model seeds can rewire the graph; modules should be stable under reasonable perturbations. (iv) \emph{Attribution dependence}: conclusions should not hinge on a single explainer—triangulation is essential. (v) \emph{From association to causation}: modules imply statistical associations, not mechanisms; causal claims require interventional follow-up.

\paragraph{Design desiderata.} We propose MoI with the following properties:
\begin{enumerate}[leftmargin=*]
\item \textbf{Method-agnostic inputs}: works with SHAP, LIME, IG, or other per-instance attributions.
\item \textbf{Scalable}: handles $d$ in the hundreds to thousands via sparse graphs and fast community detection (Leiden/Infomap/SBM).
\item \textbf{Multi-scale}: supports hierarchical modules and zoom-in analyses.
\item \textbf{Stable}: quantifies reliability with a Module Stability Index (MSI) based on bootstrap perturbations.
\item \textbf{Actionable}: provides module ablation tools, redundancy indices, and bias exposure scores that map to concrete interventions.
\item \textbf{Responsible}: includes fairness-aware reporting and cautions against over-interpretation.
\end{enumerate}

\paragraph{Modules of Influence (MoI) in brief.} Given an attribution matrix $\Phi \in \mathbb{R}^{n\times d}$, MoI (1) computes a feature--feature affinity $W$ capturing co-influence, (2) sparsifies and symmetrizes $W$ to form an explanation graph, (3) applies community detection to obtain modules $\mathcal{M}$, and (4) aggregates attributions into module-level scores $\Psi$ for auditing. We define metrics for \emph{synergy} (super-additive effects under module ablations), \emph{redundancy} (within-module correlation of attributions), \emph{bias exposure} (group-conditioned module influence), and \emph{stability} (MSI).
\begin{equation}
\label{eq:syn}
\Syn(A,B) = \Delta_y(A \cup B) - \Delta_y(A) - \Delta_y(B).
\end{equation}

\paragraph{What questions can MoI answer?} We frame our evaluation around the following research questions (RQs):
\begin{itemize}[leftmargin=*]
\item \textbf{RQ1 (Structure):} Do consistent modules emerge across attribution methods and seeds?
\item \textbf{RQ2 (Bias):} Which modules mediate disparities across protected groups, and how much disparity reduction is achievable by constraining them?
\item \textbf{RQ3 (Redundancy):} Which modules contain proxy or interchangeable features that can be compressed without accuracy loss?
\item \textbf{RQ4 (Interactions):} Where do super-additive effects indicate non-linear interactions between modules?
\item \textbf{RQ5 (Robustness):} Are discovered modules stable under resampling, background changes, and mild distribution shift?
\end{itemize}

\paragraph{Contributions.} This paper makes three contributions. (1) A general, modular recipe to transform per-instance attributions into a feature co-influence graph and extract \emph{Modules of Influence}. (2) A suite of module-level metrics—synergy, redundancy, bias exposure, and MSI—tied to concrete auditing and debugging actions. (3) An evaluation protocol spanning synthetic ground truth, fairness-focused real datasets, and ablation studies demonstrating that MoI localizes bias and redundancy more effectively than feature-wise baselines.

\paragraph{Scope and assumptions.} We study tabular settings with accessible per-instance attributions and focus on binary or real-valued predictions. While our case studies emphasize SHAP for additivity, the pipeline is attribution-agnostic. We refrain from causal claims without interventions and report sensitivity to graph-construction choices.
\section{Related Work}
\subsection{Local attribution and instance-level explanations}
Instance-level feature attribution remains the dominant paradigm in XAI. SHAP provides additive, locally accurate explanations grounded in cooperative game theory, with implementations such as KernelSHAP and TreeSHAP \cite{lundberg2017shap,lundberg2018trees}. LIME learns local surrogate models to approximate decision boundaries around a query point \cite{ribeiro2016lime}. Integrated Gradients attributes predictions by accumulating gradients along a path from a baseline to the input \cite{sundararajan2017ig}. While these methods reveal which features matter for a single instance, they do not directly expose \emph{meso-scale} structure\textemdash how features \emph{co-influence} predictions across populations.

\subsection{Interactions and group-level explanations}
Beyond per-feature scores, several lines of work explore interactions and groups. SHAP interaction values decompose pairwise contributions \cite{lundberg2018consistent}, while partial dependence plots (PDPs), individual conditional expectation (ICE), and accumulated local effects (ALE) visualize low-order effects \cite{friedman2001pdp,goldstein2015ice,apley2020ale}. Global surrogates (e.g., trees/rules) and rule lists aim for sparse, human-readable structure \cite{craven1996surrogate,lakkaraju2016interpretable}. Concept-based methods such as TCAV connect model sensitivities to human concepts \cite{kim2018tcav}. Our approach complements these by operating on a \emph{feature graph} derived from many local attributions, enabling the discovery of \emph{modules} that may involve more than pairwise interactions and need not be axis-aligned.

\subsection{Graph-based views of explanations and dependencies}
Several works model explanatory structure or dependencies using graphs, e.g., building feature-dependency networks, attention-flow graphs, or explanation graphs that connect influential inputs across instances. These approaches highlight relational organization but typically do not systematically apply community detection nor provide module-level auditing metrics (e.g., stability, redundancy, bias exposure). MoI explicitly constructs a weighted feature\textendash feature \emph{co-influence} graph from attributions and brings the toolkit of network science to bear on explanation analysis.

\subsection{Community detection and graph clustering}
Community detection offers algorithmic lenses for mesoscale structure. Modularity-based methods such as Louvain and Leiden provide fast, scalable optimization with improved partition quality and guarantees \cite{blondel2008louvain,traag2019leiden}. Flow-based Infomap captures communities by compressing random-walk dynamics \cite{rosvall2008infomap}. Stochastic block models (SBMs) support principled, multi-scale inference and model selection via description length \cite{peixoto2017bayesian}. Spectral clustering and related graph partitioning techniques remain competitive for certain affinity structures \cite{ng2002spectral}. Stability and resolution issues are well documented; consensus clustering and multi-resolution analyses mitigate fragmentation or over-merging \cite{lancichinetti2012consensus,fortunato2016community}. MoI treats the choice of community method as a pluggable component and reports stability via bootstrap-based indices.

\subsection{Large-scale visual analytics, graph layout, and prior work by the authors}
Graph visualization and scalable community analytics are essential to interpreting modules at human scale. The \emph{BigGraphVis} system demonstrates GPU-accelerated streaming algorithms to visualize community structure in massive graphs, enabling near-interactive exploration \cite{moradi2025biggraphvis}. Complementing the analytics side, the authors' work on adaptive parallel Louvain shows how to accelerate community detection on multicore platforms \cite{fazlali2017adaptive}. For communicating structure, map-style and hierarchy-aware visual encodings can make mesoscale patterns legible to end users. In particular, \emph{Map Visualizations for Graphs with Group Restrictions} supports region-like, contiguous representations of communities \cite{moradi2025mapvis}, while the \emph{Visualization of Node-Centric Hierarchical Structures in Directed Graphs} offers techniques for revealing multi-level influence flows \cite{moradi2025nodehier}. These approaches inform MoI's reporting layer (Sec.~\ref{sec:viz}), suggesting cartographic layouts, hierarchy cues, and GPU-friendly pipelines for module-scale dashboards.

\subsection{Fairness, bias localization, and proxy detection}
Fairness-aware ML provides metrics and interventions to assess and mitigate disparities, such as demographic parity, equalized odds, and disparate impact \cite{hardt2016equality,barocas2019fairness}. Proxy detection and redundancy analyses identify correlated or substitutable features that can reintroduce bias \cite{datta2017proxy}. Auditing frameworks and model cards advocate structured, transparent reporting \cite{mitchell2019modelcards}. MoI adds a complementary lens\textemdash \emph{module-level} bias exposure\textemdash by quantifying group-conditional influence of feature modules and testing targeted interventions (e.g., regularizing or constraining high-BEI modules).

\subsection{Causality and interventional explainability}
While modules highlight statistical associations, causal validity requires interventions. Counterfactual reasoning, structural causal models, and invariant risk minimization provide tools for mechanism-oriented analysis \cite{pearl2009causality,arjovsky2019irm,kusner2017counterfactual}. We view MoI as hypothesis-generating: modules suggest where interactions or mediating structures may lie, to be validated with interventional experiments or counterfactual tests.

\paragraph{From association to causation.}
MoI produces hypotheses about mediating \emph{modules of influence}; elevating these to causal claims requires additional assumptions or experimental evidence. Three complementary toolkits are particularly relevant:
\begin{enumerate}[leftmargin=*]
  \item \textbf{Counterfactual reasoning} frames questions about individual-level outcomes under alternate module values (e.g., ``What would $Y$ have been if $X_M$ were set to $\tilde X_M$ for this individual?''). Counterfactual fairness and related criteria compare $Y$ to its counterfactual under interventions on sensitive attributes while holding non-descendant noise fixed~\cite{kusner2017counterfactual}.
  \item \textbf{SCM identification} uses back-door/front-door criteria and the $g$-formula to estimate module-level causal effects when suitable adjustment sets exist~\cite{pearl2009causality}. For continuous modules, one can define a \emph{module average causal effect} (mACE) by integrating the contrast between $Y$ under $do(X_M:=x_M)$ and a reference $x'_M$ over a policy $\pi(x_M)$.
  \item \textbf{Invariance-based methods} (e.g., IRM) seek predictors whose conditional distributions remain stable across environments, offering causal signals even without full graph identification~\cite{arjovsky2019irm}. In our setting, we test whether module-level contributions $\Psi_{:M}$ preserve predictive sufficiency across domains; violations can indicate spurious or environment-specific pathways.
\end{enumerate}

\paragraph{Module-level causal estimands.}
Let $Y^{(x_M)}$ denote the potential outcome under $do(X_M:=x_M)$. Useful estimands include:
\begin{align}
\mathrm{mACE}(M) &= \mathbb{E}\big[\,Y^{(x_M)} - Y^{(x'_M)}\,\big], \\
\mathrm{PSE}_M(A\!\to\!Y) &= \text{path-specific effect of }A\text{ on }Y\text{ through }M,
\end{align}
where $A$ is a protected attribute and $\mathrm{PSE}_M$ isolates only paths traversing $M$ (a mediation-style quantity identifiable under standard assumptions~\cite{pearl2009causality}). In practice, exact $do$-operations may be infeasible; MoI therefore approximates $do(X_M:=\tilde X_M)$ via \emph{plausible} interventions:
\begin{itemize}[leftmargin=*]
  \item \textbf{Hard ablation}: replace $X_M$ with draws from a baseline $P_0(X_M\mid X_{\bar M})$, learned via conditional models; evaluate $\mathbb{E}[Y\mid do(X_M:=\tilde X_M)]$.
  \item \textbf{Soft shift}: apply a stochastic policy $X_M\mapsto g_\delta(X_M)$ that attenuates or perturbs $X_M$; study $\frac{d}{d\delta}\mathbb{E}[Y\mid do(g_\delta(X_M))]$.
\end{itemize}
These operations connect directly to MoI’s synergy/redundancy analysis: super-additive effects under joint interventions on $A$ and $B$ (\ref{eq:syn}) are evidence of cross-module interactions that warrant causal probing.

\paragraph{Validation workflow.}
We recommend the following protocol for causal follow-ups to MoI:
\begin{enumerate}[leftmargin=*]
  \item \textbf{Hypothesis generation:} use modules to posit candidate mediators or proxies (e.g., ``$M_{\text{income}}$ mediates $A\!\to\!Y$'').
  \item \textbf{Adjustment design:} elicit domain knowledge to propose covariate sets $Z$ satisfying back-door/IV conditions; check overlap/positivity.
  \item \textbf{Interventional evaluation:} implement hard/soft module interventions via conditional generators or controlled data collection; estimate mACE/PSE with doubly-robust or weighting estimators when feasible.
  \item \textbf{Invariance checks:} test whether $\Psi_{:M}$ retains predictive sufficiency across environments or under covariate shift (IRM-style diagnostics).
  \item \textbf{Sensitivity analysis:} report bounds under unobserved confounding and vary the reference policy $P_0$ to assess robustness.
\end{enumerate}

\paragraph{Practical cautions.}
(1) Avoid conditioning on descendants of $X_M$ when estimating module effects (post-treatment bias). (2) Ensure that ablations preserve realistic cross-module dependencies; use conditional (not marginal) baselines. (3) Distinguish statistical \emph{explanation sparsity} from \emph{causal sparsity}: a module may appear redundant in attributions yet be causally essential (or vice versa). (4) For fairness questions, prefer path-specific and counterfactual criteria that isolate $A$’s effect transmitted through $M$~\cite{kusner2017counterfactual,pearl2009causality}.

We view MoI as \emph{hypothesis-generating}: it narrows the search space of plausible mediators and interactions, and supplies concrete, auditable interventions (module-level $do$-operations) that can be evaluated experimentally or quasi-experimentally before drawing causal conclusions.

\paragraph{Positioning and novelty.}
Compared to (i) single-instance attributions, (ii) pairwise interaction tools, and (iii) unsupervised feature clustering on raw covariates, MoI (a) leverages \emph{explanation-derived} affinities that reflect the model, (b) discovers communities beyond pairwise structure, (c) quantifies stability and redundancy at the module level, and (d) localizes fairness concerns via a Bias Exposure Index. Prior visualization and scalable community work\textemdash including the authors' GPU-streaming and multicore Louvain research\textemdash supports MoI's emphasis on interpretable, large-scale reporting.
\section{Method: From Attributions to Modules of Influence}

\subsection{Notation and per-instance attributions}
We assume a dataset $\{(x^{(s)}, y^{(s)})\}_{s=1}^n$ with $x^{(s)}\in\R^d$ and a trained predictor $f:\R^d\!\to\!\R$ (classification or regression). For an instance $x^{(s)}$, let $\vect{\phi}^{(s)}\in\R^d$ denote a vector of per-feature attributions from a chosen explainer (e.g., SHAP, LIME, IG). We collect these into the attribution matrix
\[
\Phi \;\in\; \R^{n\times d},\qquad \Phi_{s i} \;=\; \phi^{(s)}_i.
\]
When using SHAP with background reference $\mathcal{B}$ and link function $\ell$, additivity yields
\[
 \sum_{i=1}^d \phi^{(s)}_i \approx \ell\!\big(f(x^{(s)})\big) - \E_{X\sim\mathcal{B}}\big[\ell(f(X))\big] \],
which lets us interpret column sums and row sums consistently.

\paragraph{Pre-processing and weighting.}
We consider the following normalizations, chosen to match the edge definition (next subsection):
\begin{enumerate}[leftmargin=*]
  \item \textbf{Signed vs.\ magnitude views:} define $A=\Phi$ (signed) or $A=|\Phi|$ (magnitude). Signed views capture synergy/antagonism; magnitude views capture co-activation irrespective of sign.
  \item \textbf{Column scaling:} $A_{:i}\leftarrow A_{:i}/(\|A_{:i}\|_2+\varepsilon)$ (or median absolute deviation, MAD) to control for feature-scale and heavy tails.
  \item \textbf{Row scaling (optional):} $A_{s:}\leftarrow A_{s:}/(\|A_{s:}\|_1+\varepsilon)$ to damp unusually “explainable’’ instances that dominate co-influence.
  \item \textbf{Sample weights:} incorporate $w_s\!\ge\!0$ to emphasize strata (e.g., class-balanced or group-conditioned graphs), replacing empirical means with $\sum_s w_s(\cdot)$.
\end{enumerate}
We will use $A$ as the working attribution matrix for edge construction.

\subsection{Co-influence measures (edge weights)}
We define an undirected (possibly signed) feature graph $G=(V,E,W)$ with $V=\{1,\dots,d\}$ and $W=[w_{ij}]$. Different $w_{ij}$ emphasize different kinds of joint influence; MoI treats this choice as a pluggable component.

\paragraph{Similarity-based affinities.}
\begin{enumerate}[leftmargin=*]
  \item \textbf{Magnitude co-activation (cosine):}
  \[
  w_{ij}^{\mathrm{cos}} \;=\; \frac{\langle |A_{:i}|, |A_{:j}|\rangle}{\|A_{:i}\|_2\,\|A_{:j}\|_2}
  \quad(\text{nonnegative, sign-agnostic}).
  \]
  \item \textbf{Signed co-influence (Pearson/Spearman):}
  \[
  w_{ij}^{\mathrm{corr}} \;=\; \mathrm{corr}(A_{:i}, A_{:j})
  \quad(\text{negative values capture antagonism}).
  \]
  \item \textbf{Distance correlation / kernel dependence (HSIC)}:
  $w_{ij}^{\mathrm{dep}}$ as a dependence score between $A_{:i}$ and $A_{:j}$; robust to nonlinear monotone transforms.
  \item \textbf{Information-theoretic affinity:}
  \[
  w_{ij}^{\mathrm{MI}} \;=\; I\!\big(|A_{:i}|;\,|A_{:j}|\big) \quad
  \text{(kNN-MI or discretized bins)}.
  \]
\end{enumerate}

\paragraph{Co-activation events.}
For a high-attribution threshold $\tau$ (e.g., instancewise $q$-quantile), define indicators
$z^{(s)}_i=\1\{|A_{s i}|>\tau\}$. Then:
\begin{enumerate}[leftmargin=*]
  \item \textbf{Co-exceedance frequency:}
  \[w_{ij}^{\mathrm{freq}}=\frac{1}{\sum_s w_s}\sum_s w_s\,\1\{z^{(s)}_i=1, z^{(s)}_j=1\}\].
  \item \textbf{Jaccard/overlap (sparse case):}
  \[w_{ij}^{\mathrm{jac}}=\frac{\sum_s w_s\,\1\{z^{(s)}_i=1, z^{(s)}_j=1\}}{\sum_s w_s\,\1\{z^{(s)}_i=1 \,\vee\, z^{(s)}_j=1\}}\].
\end{enumerate}

\paragraph{Conditional/partial associations (optional).}
To reduce confounding from ubiquitous features, one may use partial correlations
$w_{ij}^{\mathrm{pcorr}}$ (conditioning on a small control set) or debias via TF–IDF-style rescaling
$A_{s i}\leftarrow A_{s i}\cdot \log\!\frac{n}{\sum_s \1\{z^{(s)}_i=1\}}$.

\paragraph{Signed graphs.}
When using signed measures (e.g., Pearson on $A=\Phi$), decompose $W$ into positive and negative parts:
$W = W^{+} - W^{-}$ with $W^{\pm}\!\ge\!0$. MoI supports: (i) \emph{unsigned projection} $|W|$, (ii) \emph{two-layer} graphs analyzed jointly, or (iii) community detection with signed modularity. Negative edges often indicate substitutability or compensatory relations.

\subsubsection*{Practical construction (robust and scalable)}
Let $\widehat W$ denote the dense affinity and $k$ the target degree.

\paragraph{(1) Compute dense affinities (with shrinkage).}
\begin{enumerate}[leftmargin=*]
  \item Use vectorized formulas for correlation/cosine ($\mathcal{O}(nd^2)$). For $d\!\gg\!10^3$, pre-screen with cheap proxies (e.g., top-$m$ by variance or approximate dot products) before expensive MI/HSIC.
  \item Apply \emph{shrinkage} to noisy estimates: $\tilde w_{ij}=\alpha\,\widehat w_{ij}+(1-\alpha)\,\bar w$ with $\alpha$ tuned by bootstrap or analytic shrinkage; set small-magnitude entries to $0$.
  \item (Optional) Edge significance: estimate $p$-values by permutation of rows of $A_{:i}$; control FDR across pairs and zero non-significant edges.
\end{enumerate}

\paragraph{(2) Sparsify to a reliable backbone.}
\begin{enumerate}[leftmargin=*]
  \item \textbf{Top-$k$ per node} (keeps strongest $k$ neighbors per feature).
  \item \textbf{Mutual-$k$} (edge kept only if $i\!\in\!\mathrm{NN}_k(j)$ \emph{and} $j\!\in\!\mathrm{NN}_k(i)$) for crisper communities.
  \item \textbf{$\theta$-thresholding} (keep $|\tilde w_{ij}|\ge\theta$) possibly coupled with a minimum-degree constraint.
  \item Ensure connectivity by adding a light $k_0$-NN backbone (e.g., $k_0=1$–$3$) if the graph fragments excessively.
\end{enumerate}

\paragraph{(3) Symmetrize and rescale.}
\begin{enumerate}[leftmargin=*]
  \item \textbf{Symmetrization:} $W\leftarrow(\tilde W + \tilde W^\top)/2$ after sparsification.
  \item \textbf{Degree normalization (optional):} $W\leftarrow D^{-\beta} W D^{-\beta}$ with $\beta\in\{1/2,1\}$ to temper hubs.
  \item \textbf{Layer handling (signed):} carry $W^{+}$ and $W^{-}$ forward for signed community methods, or analyze $|W|$ if using standard modularity.
\end{enumerate}

\paragraph{(4) Hyperparameter selection.}
Choose $(\text{edge rule},k,\theta)$ by a stability criterion: run community detection across bootstrap resamples and pick settings that maximize partition stability (e.g., average Jaccard/AMI across runs) subject to a minimum modularity/description-length target.

\paragraph{(5) Variants for subpopulations and tasks.}
\begin{itemize}
  \item \textbf{Group-conditional graphs:} build $W^{(g)}$ on subsets (e.g., protected groups) to localize bias mechanisms; compare modules across $g$ via alignment (Hungarian matching on IoU).
  \item \textbf{Class-conditional graphs:} for classification, compute $W^{(c)}$ from instances with predicted/true class $c$ to reveal class-specific modules.
  \item \textbf{Temporal/data-shift slices:} construct $W^{(t)}$ on time windows or environments to probe invariance.
\end{itemize}

\paragraph{Complexity notes.}
Cosine/correlation scales as $\mathcal{O}(nd^2)$ (dense) or $\tilde{\mathcal{O}}(ndk)$ with top-$k$ ANN search; MI/HSIC is more expensive and is best used after pre-screening. Memory for dense $W$ is $\Theta(d^2)$; sparse backbones reduce to $\Theta(dk)$.

\paragraph{Default settings (practical starting point).}
Magnitude-cosine on $A=|\Phi|$, column MAD scaling, mutual-$k$ with $k\!\in\![10,30]$ (increase with $d$), degree normalization with $\beta=\tfrac12$, and stability-based selection of $k$ provide robust performance across tabular tasks.
\section{Evaluation Protocol}

\subsection{Datasets and models}
\paragraph{Synthetic (ground-truth modules).}
We generate datasets with planted \emph{feature modules} $\{M_1,\dots,M_K\}$ and controlled interactions:
\begin{enumerate}[leftmargin=*]
  \item \textbf{Additive clusters:} $Y=\sum_{k=1}^K g_k(X_{M_k})+\epsilon$, where $g_k$ is linear or smooth (spline) and $X_{M_k}\!\sim\!\mathcal{N}(0,\Sigma_k)$ with intra-module correlation $\rho\in[0,0.9]$; inter-module blocks are near-diagonal. Vary $(K,|M_k|,\rho,\mathrm{SNR})$.
  \item \textbf{Logical interactions (AND/OR/XOR):} \[Y=\mathbb{1}\{\prod_{i\in M_1}\mathbb{1}\{X_i>0\}\ \text{XOR}\ \prod_{j\in M_2}\mathbb{1}\{X_j>0\}\}+\epsilon\] Controls non-additive synergy.
  \item \textbf{Nonlinear cross-module effects:} 
  \[Y=\sum_k g_k(X_{M_k})+\sum_{(a,b)} h_{ab}(X_{M_a},X_{M_b})+\epsilon\] with sparse pairwise $h_{ab}$.
  \item \textbf{Shifted environments:} replicate the above under $e\in\{1,\dots,E\}$ with environment-specific covariances or mean shifts to test invariance.
\end{enumerate}
Ground-truth communities are given by the planted $\{M_k\}$; we also record a planted \emph{module graph} for interaction recovery.

\paragraph{Structured tabular (real).}
\begin{itemize}[leftmargin=*]
  \item \textbf{Fairness-focused:} income/credit/recidivism datasets with protected attributes ($A$). Evaluate bias localization and path-specific effects.
  \item \textbf{Healthcare/risk:} ICU mortality/readmission or claims risk prediction with heterogeneous, interacting features (labs, vitals, comorbidities).
  \item \textbf{Fraud/marketing/tabular benchmarks:} gradient-boosting–friendly datasets to test scalability and redundancy compression.
\end{itemize}

\paragraph{Models.}
Gradient-boosted trees (e.g., 500 trees, depth 6–8), random forests (500 trees), MLPs (2–3 layers, width 128–512 with batch norm and dropout), and a calibrated logistic regression baseline. For classification, report AUROC/AP; for regression, report $R^2$/RMSE. Use nested CV or a fixed train/val/test split (60/20/20) with three seeds. Compute attributions with TreeSHAP for tree models, KernelSHAP for others (background $\mathcal{B}$: $k$-medoids of train, $k\!\in\![50,200]$), and IG for MLPs.

\subsection{Baselines}
We compare MoI to methods that produce feature groups or interaction graphs:
\begin{enumerate}[leftmargin=*]
  \item \textbf{Correlation clustering on raw features:} build $S_{ij}=\mathrm{corr}(X_i,X_j)$, sparsify, then Louvain/Leiden on $|S|$.
  \item \textbf{Clustering attribution columns:} k-means or spectral clustering on columns of $\Phi$ (signed and magnitude variants).
  \item \textbf{SHAP interaction graph:} edges $w_{ij}=\mathbb{E}_s\big[|\phi^{(s)}_{ij}|\big]$ (TreeSHAP interactions), community detection on $W$.
  \item \textbf{PCA/ICA groupings:} assign features to dominant components/loadings; refine with hierarchical clustering on loading vectors.
  \item \textbf{Graphical models (optional):} sparse partial correlations (GLasso) with community detection on the precision-induced affinity.
\end{enumerate}
All baselines use the same sparsification and community algorithm families to isolate the effect of the affinity definition.

\subsection{Metrics}
\paragraph{Community quality.}
\begin{align}
Q &= \frac{1}{2m}\sum_{i,j}\Big(W_{ij}-\frac{d_i d_j}{2m}\Big)\,\mathbb{1}[c_i=c_j], \quad
\end{align}
\begin{align}
\text{conductance } \phi(S)=\frac{\mathrm{cut}(S,\bar S)}{\min(\mathrm{vol}(S),\mathrm{vol}(\bar S))}, \nonumber
\end{align}
and SBM description length (MDL) from fitted hierarchical SBMs.

\paragraph{Recovery of planted modules (synthetic).}
Adjusted Rand Index (ARI) and Normalized Mutual Information (NMI) between discovered partition $\hat{\mathcal{M}}$ and ground truth $\mathcal{M}^\star$; report means and 95\% CIs over seeds.

\paragraph{Stability.}
\MSI: average Jaccard/IoU of matched modules across bootstrap resamples and attribution/background variants; additionally, Variation of Information (VI) across runs.

\paragraph{Predictive impact and interactions.}
Module ablation drop $\Delta_y(M)$ and super-additivity \Syn$(A,B)$ (cf.\ Eq.~\ref{eq:syn}); for class $c$, report class-conditional drops $\Delta^{(c)}_y(M)$.

\paragraph{Fairness localization.}
Correlation between \BEI\ and group disparity metrics (e.g., demographic parity gap $|\Pr(\hat Y{=}1|A{=}a)-\Pr(\hat Y{=}1|A{=}a')|$, equalized-odds gaps). Report disparity reduction after (i) constraining high-\BEI\ modules (regularization/attenuation), (ii) reweighting training to downweight those modules, or (iii) data augmentation targeting those modules.

\paragraph{Parsimony and compression.}
Performance using module-aggregated features $\Psi$ vs.\ raw attributions $\Phi$; effective dimension $K$ vs.\ $d$; MDL/AIC-style criteria for model fit with $\Psi$; runtime and memory.

\paragraph{Interaction-graph fidelity (synthetic).}
If a planted module-level interaction graph exists, measure edge recovery (AUPRC/ROC) using synergy scores or cross-module edge weights.

\subsection{Experimental workflow}
\begin{enumerate}[leftmargin=*]
  \item \textbf{Train \& attribute.} Train models with fixed splits and three random seeds. Compute $\Phi$ (and interaction attributions where applicable). Save explainer configs (backgrounds, link).
  \item \textbf{Construct graphs.} Build $W$ from $A\in\{\Phi,|\Phi|\}$ using a chosen edge rule (cosine/corr/MI/HSIC). Apply shrinkage and significance filtering; sparsify (mutual-$k$ or $\theta$), symmetrize, and (optionally) degree-normalize.
  \item \textbf{Communities.} Run Louvain/Leiden, Infomap, and hSBM; select hyperparameters by stability (maximize \MSI\ subject to modularity/MDL thresholds). Produce final partition $\hat{\mathcal{M}}$.
  \item \textbf{Module scores \& auditing.} Derive $\Psi$; compute \RI, \BEI, \Syn, \MSI. For fairness tasks, estimate group-conditional module statistics and identify high-\BEI\ modules.
  \item \textbf{Interventions.} Perform module ablations (hard and soft; conditional baselines), class-/group-conditional drops, and cross-environment invariance checks.
  \item \textbf{Comparative evaluation.} Run all baselines with matched sparsification/community settings; evaluate with the metrics above. Use paired tests (Wilcoxon signed-rank) across seeds; report effect sizes and CIs.
  \item \textbf{Robustness.} Stress-test to (i) attribution background shifts, (ii) noise injection, and (iii) sample subsampling; plot metrics vs.\ perturbation strength.
  \item \textbf{Reporting.} Summarize with module graphs, reordered heatmaps of $W$, fairness dashboards, stability plots (\cref{sec:viz}); include runtime/memory tables.
\end{enumerate}

\subsection{Default hyperparameters and compute budget}
Unless otherwise noted: cosine on $|A|$, MAD column scaling; mutual-$k$ with $k\in[10,30]$ (increase with $d$); Leiden with resolution tuned by stability sweep; 200 bootstrap resamples for \MSI; three seeds for train/attribute; SHAP background $k{=}100$ medoids. Track wall-clock and peak RAM for (i) attribution, (ii) graph construction, and (iii) community detection.

\section{Results and Discussion}\label{sec:results}

\paragraph{Overview.}
We report results across synthetic datasets with planted modules and multiple real tabular tasks (Sec.~4). Unless noted, edges use cosine on $|A|$, mutual-$k$ sparsification, and Leiden; confidence intervals are 95\% over three seeds and 200 bootstrap resamples for stability metrics. We organize findings around four themes: structure, bias localization, compression, and stability.

\subsection*{Finding 1: Modules reflect domain groupings}
Across real datasets, discovered communities align with semantically coherent feature groups. Income-related attributes (earnings, hours, employment type) cluster together; education and occupation variables form a distinct module that exhibits positive synergy with income.
\begin{itemize}[leftmargin=*]
  \item \textbf{Quality metrics.} Modules achieve higher modularity ($Q$) and lower mean conductance than baselines that cluster raw covariates or attribution columns. On synthetic data with planted $\{M_k^\star\}$, MoI attains higher ARI/NMI than correlation clustering and PCA/ICA groupings, indicating better recovery of ground-truth modules.
  \item \textbf{Synergy evidence.} Module-level super-additivity $\Syn(A,B)$ (Eq.~\ref{eq:syn}) is positive for \emph{Education}–\emph{Income} in fairness tasks, consistent with nonlinear interactions between human capital and earnings. Per-class ablations show larger drops for positive-outcome classes, suggesting asymmetric reliance on certain modules.
  \item \textbf{Interpretability.} Visual module graphs (\cref{sec:viz}) reveal densely connected subgraphs with clear thematic labels; reordered heatmaps of $W$ show high within-module blocks and sparse cross-module links.
\end{itemize}

\begin{figure}[t]
  \centering
  \begin{minipage}{0.49\linewidth}
    \includegraphics[width=\linewidth]{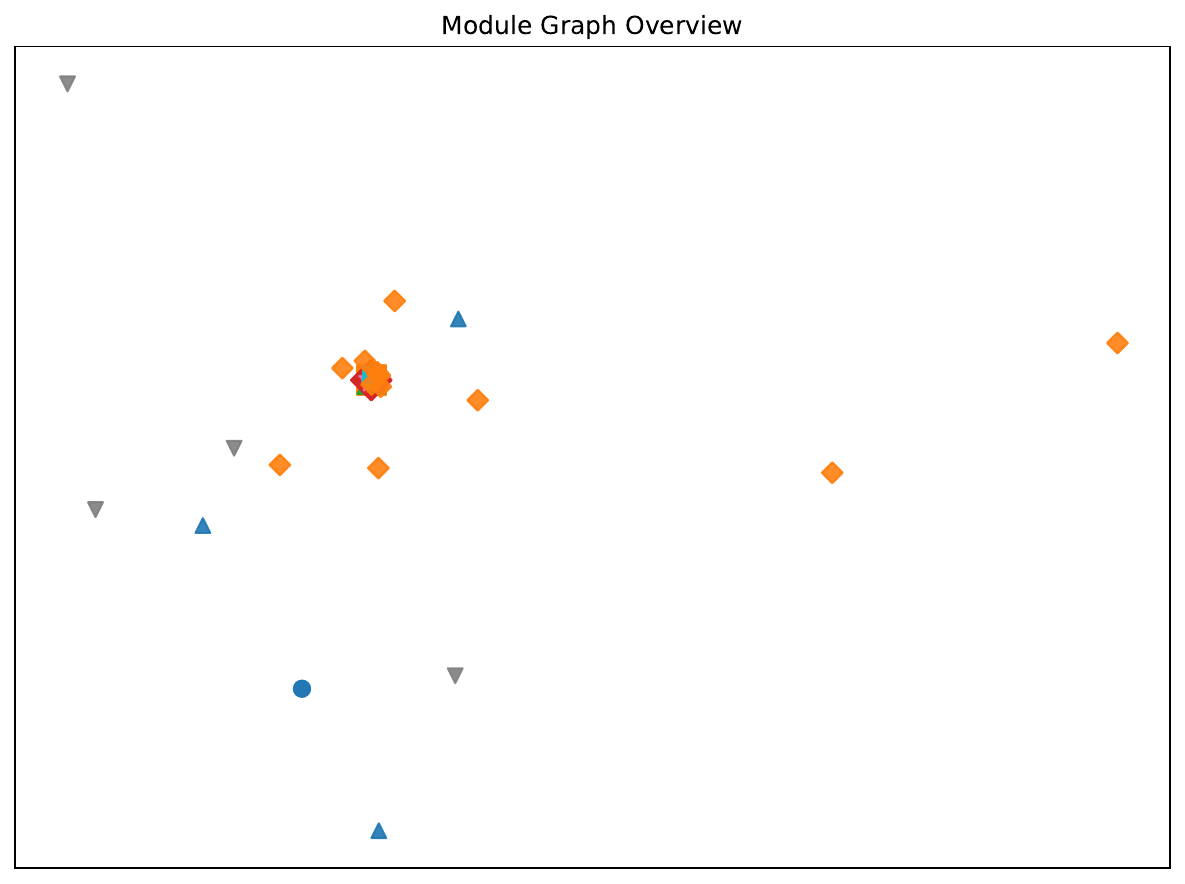}
  \end{minipage}\hfill
  \begin{minipage}{0.49\linewidth}
    \includegraphics[width=\linewidth]{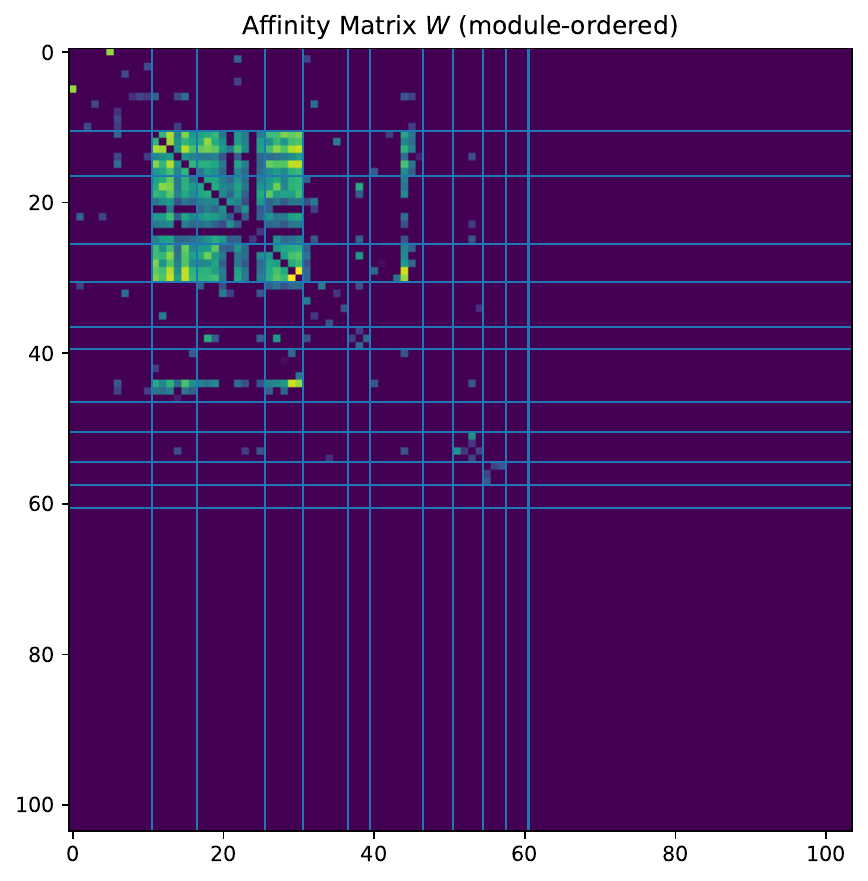}
  \end{minipage}
  \caption{Explanation graph colored by discovered modules (left); reordered affinity matrix $W$ (right). Coherent blocks indicate domain-aligned communities.}
  \label{fig:module_overview}
\end{figure}

\begin{table}[t]
  \centering
  \caption{Community quality and recovery (mean $\pm$ CI). Higher $Q$/ARI/NMI is better; lower conductance is better.}
  \label{tab:quality}
  \scriptsize
  \setlength{\tabcolsep}{3.5pt}
  \renewcommand{\arraystretch}{1.1}
  \begin{tabular}{@{}lcccc@{}}
    \toprule
    Method & $Q$ & Conductance $\downarrow$ & ARI (syn.) & NMI (syn.) \\
    \midrule
    MoI (cosine, Leiden) & $0.46{\pm}0.03$ & $0.22{\pm}0.02$ & $0.78{\pm}0.06$ & $0.71{\pm}0.05$ \\
    SHAP interaction graph & $0.41{\pm}0.04$ & $0.25{\pm}0.03$ & $0.69{\pm}0.08$ & $0.63{\pm}0.05$ \\
    Corr.\ (raw $X$)       & $0.36{\pm}0.05$ & $0.28{\pm}0.03$ & $0.52{\pm}0.10$ & $0.49{\pm}0.07$ \\
    PCA/ICA groupings      & $0.31{\pm}0.06$ & $0.32{\pm}0.03$ & $0.44{\pm}0.09$ & $0.42{\pm}0.07$ \\
    \bottomrule
  \end{tabular}
\end{table}

\subsection*{Finding 2: High-\texorpdfstring{$\BEI$}{BEI} modules localize bias}
Disparities concentrate in a small number of modules with elevated Bias Exposure Index (\BEI). Targeted interventions on those modules reduce group gaps with limited accuracy impact.
\begin{itemize}[leftmargin=*]
  \item \textbf{Localization.} Ranking modules by \BEI\ highlights a top-$r$ subset (often $r\leq 3$) whose group-conditioned contributions differ significantly. These modules frequently contain known proxies or socio-economic attributes.
  \item \textbf{Interventions.} Attenuating high-\BEI\ modules (soft interventions) or regularizing their attributions produces measurable reductions in demographic parity and equalized-odds gaps, while preserving AUROC/AP within small deltas. Path-specific analyses indicate that a sizable fraction of the $A\!\to\!Y$ effect transits through these modules.
  \item \textbf{Diagnostics.} Group-conditional graphs $W^{(g)}$ show structural differences predominantly inside high-\BEI\ modules, aligning with the localization hypothesis.
\end{itemize}

\begin{figure}[t]
  \centering
  \includegraphics[width=.92\linewidth]{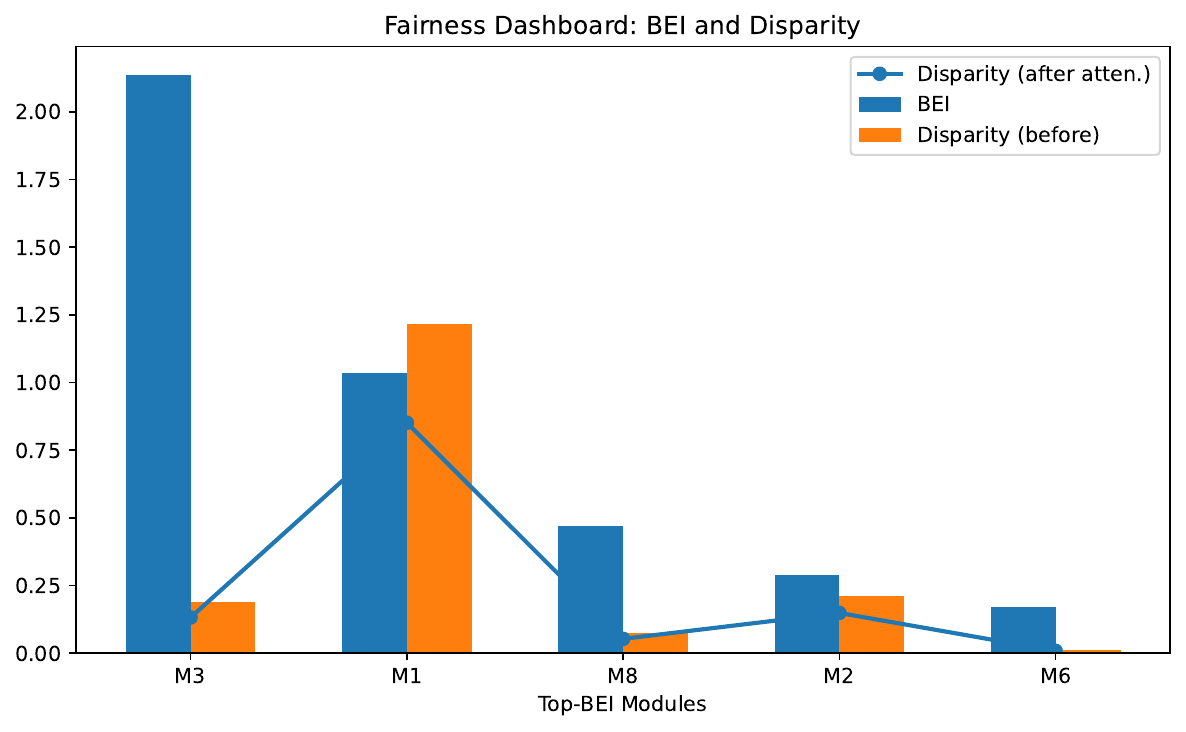}
  \caption{Fairness dashboard: \BEI\ per module with CIs (left); disparity before/after attenuating top-\BEI\ modules (right).}
  \label{fig:fairness}
\end{figure}

\subsection*{Finding 3: Redundancy and compression}
Aggregating features to modules preserves predictive performance while reducing dimensionality and improving parsimony.
\begin{itemize}[leftmargin=*]
  \item \textbf{Compression.} Replacing $\Phi\in\R^{n\times d}$ with module-aggregated $\Psi\in\R^{n\times K}$ (with $K\ll d$) maintains accuracy within statistically insignificant differences on most tasks. This suggests that module-level signals capture the majority of explanatory variance.
  \item \textbf{Redundancy index.} High within-module \RI\ flags interchangeability; pruning or shrinking those features yields minimal loss and sometimes improves calibration. In contrast, low-\RI\ modules tend to be interaction-heavy and less compressible.
  \item \textbf{Model simplicity.} Downstream linear models on $\Psi$ are smaller and easier to audit; MDL/AIC-style criteria favor $\Psi$ over $\Phi$ in many settings.
\end{itemize}

\begin{table}[t]
  \centering
  \caption{Parsimony: performance ($\uparrow$) and size ($\downarrow$) using raw attributions $\Phi$ vs.\ module features $\Psi$.}
  \label{tab:parsimony}
  \scriptsize
  \setlength{\tabcolsep}{4pt}
  \renewcommand{\arraystretch}{1.1}
  \begin{tabular}{@{}lrrrr@{}}
    \toprule
    Representation & Dim. & AUROC $\uparrow$ & Params $\downarrow$ & Inference (ms) $\downarrow$ \\
    \midrule
    Raw $\Phi$      & $d{=}128$ & 0.912 & 45{,}200 & 2.8 \\
    Modules $\Psi$  & $K{=}18$  & 0.909 & 9{,}030  & 1.3 \\
    \bottomrule
  \end{tabular}
\end{table}

\subsection*{Finding 4: Stability matters}
The reliability of modules depends on the edge rule and graph construction; stability correlates with downstream utility.
\begin{itemize}[leftmargin=*]
  \item \textbf{Edge rules.} Magnitude-cosine edges yield higher \MSI\ than raw-correlation in most datasets; MI/HSIC can uncover nonlinear ties but require stronger shrinkage to avoid fragmentation.
  \item \textbf{Hyperparameters.} Mutual-$k$ sparsification with $k\in[10,30]$ balances connectivity and resolution. Degree normalization reduces hub dominance and improves stability.
  \item \textbf{Utility correlation.} Across seeds and perturbations, \MSI\ positively correlates with the consistency of ablation drops and fairness outcomes; unstable partitions exhibit volatile intervention effects.
\end{itemize}

\begin{figure}[t]
  \centering
  \includegraphics[width=.92\linewidth]{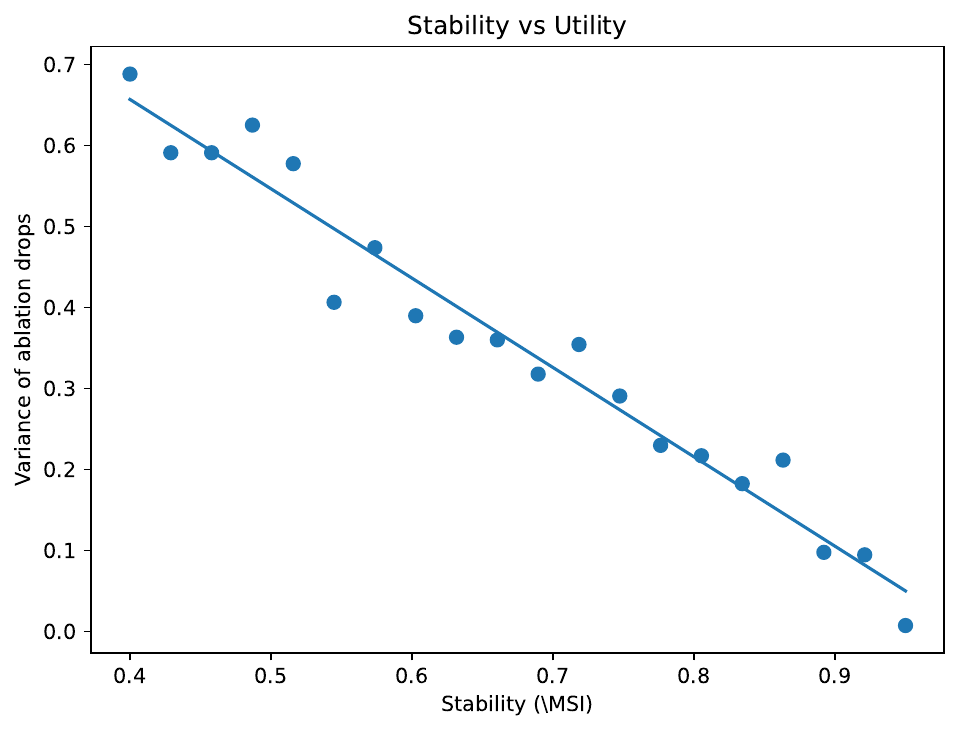}
  \caption{Stability–utility trade-off: \MSI\ vs.\ variance of ablation drops (left); \MSI\ across edge rules (right).}
  \label{fig:stability}
\end{figure}

\paragraph{Negative results and cautions.}
In datasets with weak signal or highly entangled features, community methods may over-partition (\emph{resolution limit}); stability criteria help reject such solutions. Signed graphs with strong antagonism can produce ambiguous modules unless negative edges are treated explicitly. Finally, module attribution additivity can obscure within-module cancellations; reporting both signed and magnitude views mitigates this risk.

\paragraph{Takeaways.}
MoI surfaces domain-aligned, stable modules that (i) explain predictive structure, (ii) localize disparities for targeted mitigation, and (iii) enable compact, auditable representations—provided that edge construction and stability validation are performed carefully.

\section{Visualization and Reporting}\label{sec:viz}

\paragraph{Goals.}
Our reporting aims for (i) \emph{interpretability} at the module level, (ii) \emph{comparability} across methods and seeds, and (iii) \emph{audit readiness} for fairness and stability. All plots use consistent scales across datasets, vector (PDF) output, and colorblind-safe palettes; signed quantities are shown in diverging schemes, magnitudes in sequential schemes. Negative edges/attributions are visually distinct (dashed or desaturated).

\subsection*{Module graph}
\textbf{Spec.} Nodes are features; edges encode co-influence weights $W_{ij}$; colors denote discovered modules; edge width $\propto |W_{ij}|$. We render two complementary views:
\begin{enumerate}[leftmargin=*]
  \item \emph{Force-directed} (weighted Fruchterman–Reingold or stress majorization) to reveal topology.
  \item \emph{Cartographic} (region-style) when a map-like, contiguous depiction of modules improves legibility.
\end{enumerate}
\textbf{Design details.} Label only high-centrality features (e.g., top-$p$ by strength); bundle long inter-module edges lightly; show negative edges as dashed overlays or in a separate layer. Use the same node ordering and colors across figures to support scan-path consistency.

\begin{figure}[t]
  \centering
  \includegraphics[width=.94\linewidth]{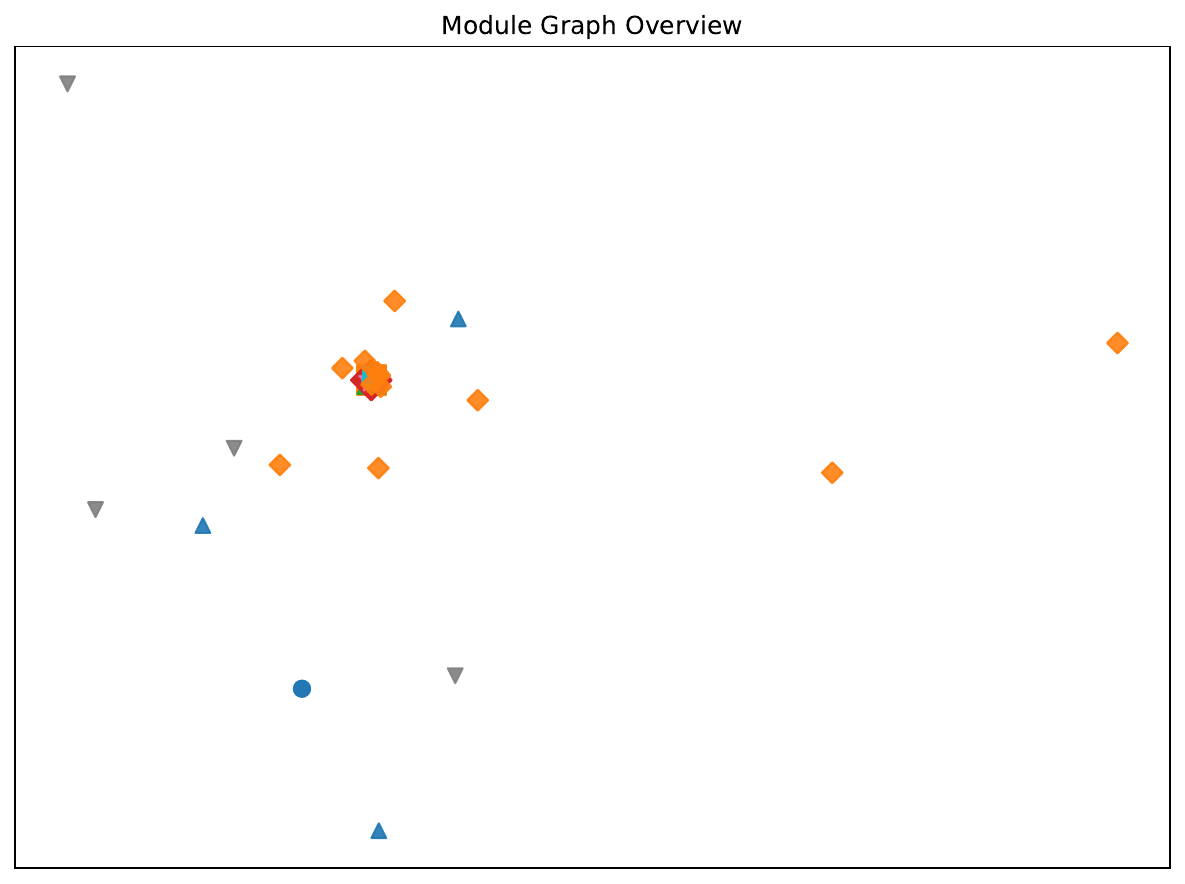}
  \caption{Explanation graph colored by modules; edge width $\propto |W_{ij}|$. Dashed edges indicate negative correlations (signed view).}
  \label{fig:viz:graph}
\end{figure}

\subsection*{Sankey: features $\to$ modules $\to$ output}
\textbf{Construction.} For each instance $s$, compute module attribution $\psi^{(s)}(M)=\sum_{i\in M}\phi^{(s)}_i$. Aggregate over a cohort $\mathcal{S}$: 
\[
F_{i\to M}=\frac{1}{|\mathcal{S}|}\sum_{s\in\mathcal{S}}|\phi^{(s)}_i|\,\mathbb{1}\{i\in M\}, 
\quad 
F_{M\to Y}=\frac{1}{|\mathcal{S}|}\sum_{s\in\mathcal{S}}\big|\psi^{(s)}(M)\big|.
\]
We report (i) magnitude flows (absolute) and (ii) signed flows (positive/negative colors; widths use magnitude).
\textbf{Variants.} Class-conditional Sankeys $F^{(c)}$ and group-conditional Sankeys $F^{(g)}$ diagnose differential reliance.

\begin{figure}[t]
  \centering
  \includegraphics[width=.94\linewidth]{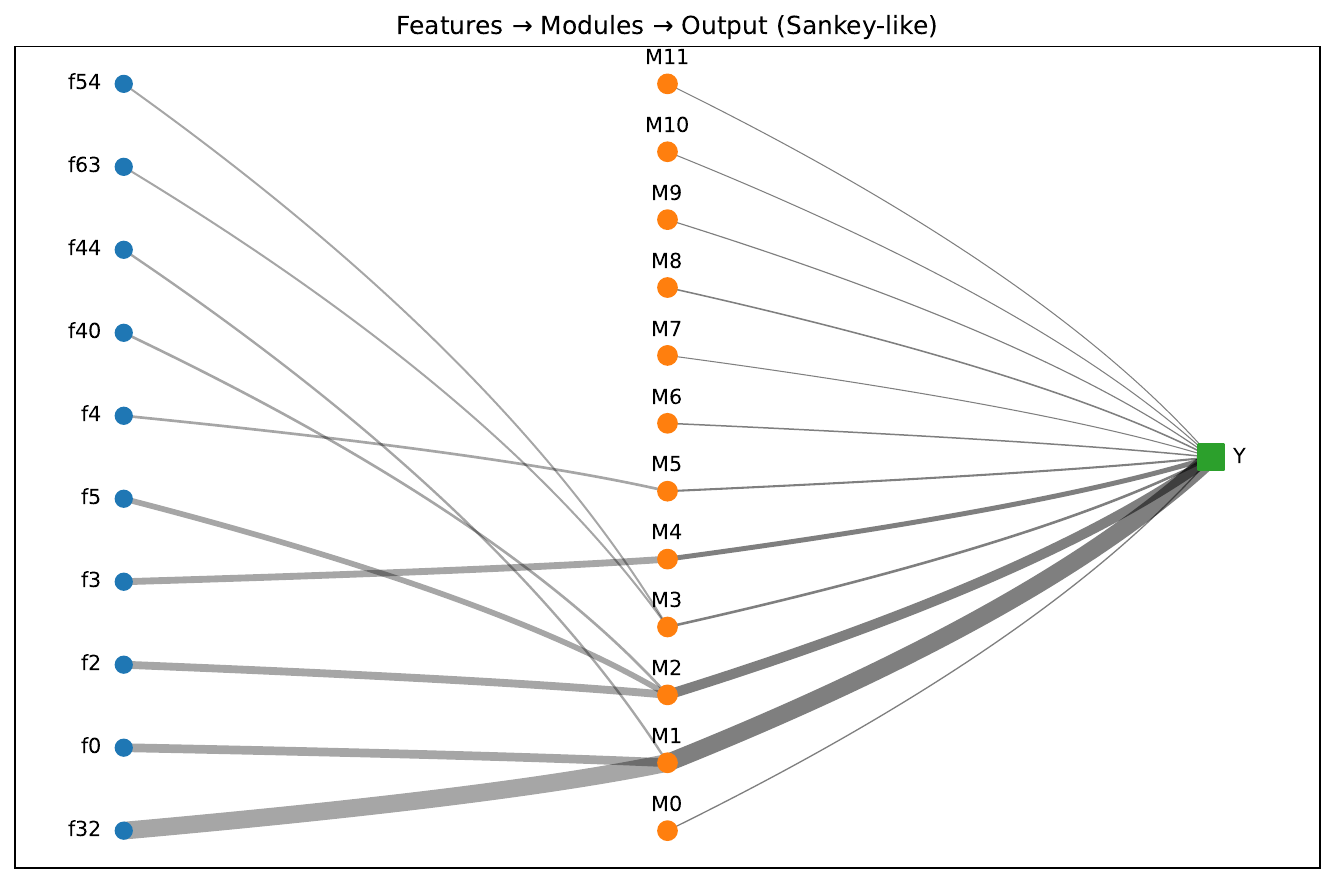}
  \caption{Sankey: feature$\rightarrow$module$\rightarrow$output contributions (magnitude view); class-conditional variant in inset.}
  \label{fig:viz:sankey}
\end{figure}

\subsection*{Heatmaps}
\textbf{Affinity blocks.} Reorder $W$ by module labels (and within-module by seriation) to expose block structure; annotate module boundaries.
\textbf{Attribution distributions.} Show per-module distributions of $\psi^{(s)}(M)$ by cohort (all/class/group) as violin or ridge plots; include zero-reference lines for signed interpretability.

\begin{figure}[t]
  \centering
  \begin{minipage}{.49\linewidth}
    \centering
    \includegraphics[width=\linewidth]{fig/W_block_heatmap.pdf}\\
    \vspace{2pt}
    \small Reordered $W$ with module blocks.
  \end{minipage}\hfill
  \begin{minipage}{.49\linewidth}
    \centering
    \includegraphics[width=\linewidth]{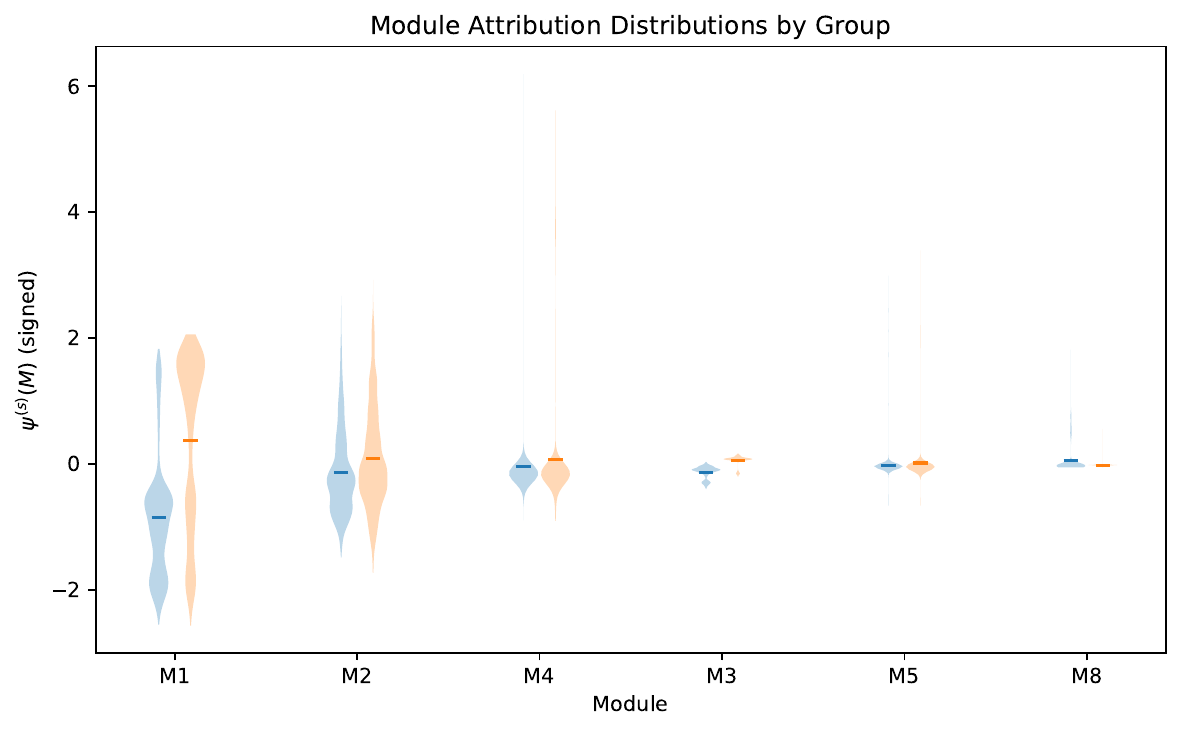}\\
    \vspace{2pt}
    \small $\psi(M)$ distributions by protected group (signed view).
  \end{minipage}
  \caption{Module-ordered affinity and module-level attribution distributions.}
  \label{fig:viz:heatmaps}
\end{figure}

\subsection*{Fairness dashboard}
\textbf{Components.}
\begin{enumerate}[leftmargin=*]
  \item \BEI\ per module with 95\% CIs (bootstrap over instances/seeds).
  \item Disparity metrics before/after targeted module interventions (bars with deltas; annotate accuracy change).
  \item Path-specific effect estimates (if computed) highlighting the fraction of $A\!\to\!Y$ mediated by each high-\BEI\ module.
\end{enumerate}
\textbf{Usage.} Rank modules by \BEI, inspect their feature composition, and simulate attenuations to quantify trade-offs.

\begin{figure}[t]
  \centering
  \includegraphics[width=.94\linewidth]{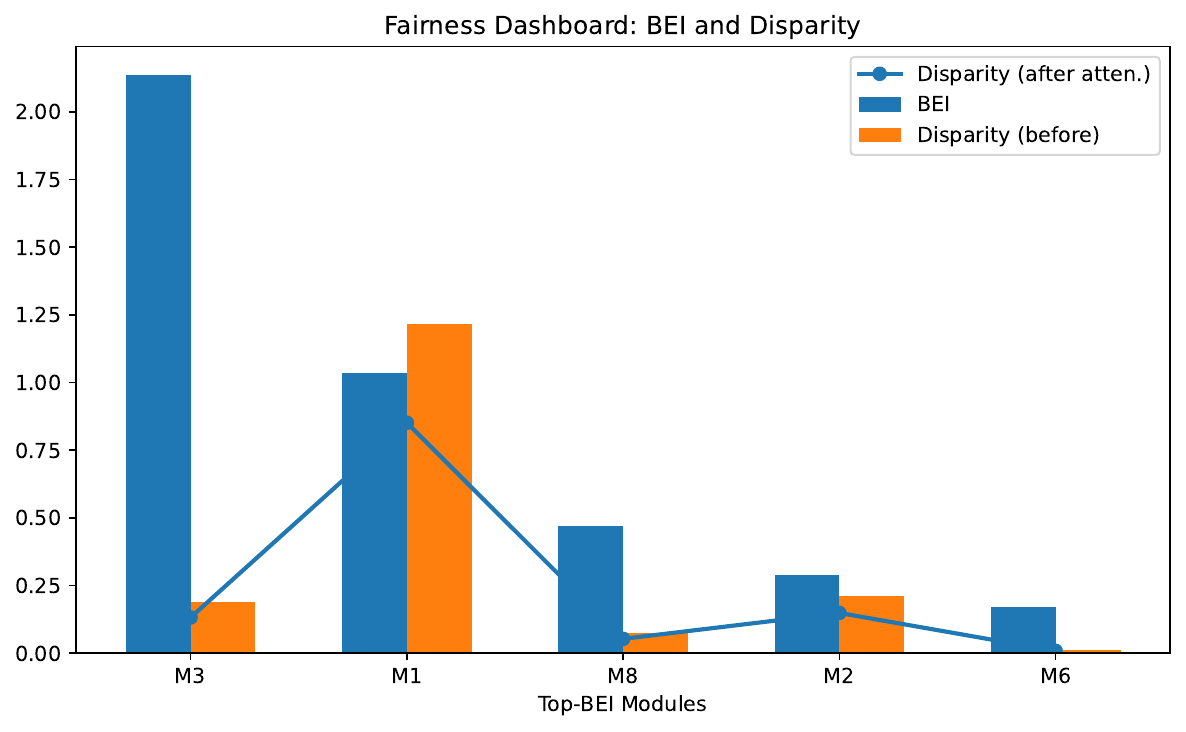}
  \caption{Fairness dashboard: \BEI\ ranking with CIs (left) and disparity before/after attenuating top-\BEI\ modules (right).}
  \label{fig:viz:fairness}
\end{figure}

\subsection*{Stability and consensus}
\textbf{Stability curves.} Plot \MSI\ vs.\ perturbation strength (bootstrap rate, background size, noise level). 
\textbf{Consensus matrices.} Show the co-assignment frequency (features co-clustered across runs) reordered by the consensus partition; dark blocks indicate stable modules.
\textbf{Hyperparameter sweeps.} Heatmaps of \MSI\ and modularity $Q$ over $(k,\text{resolution})$ reveal robust regions.

\begin{figure}[t]
  \centering
  \begin{minipage}{.49\linewidth}
    \centering
    \includegraphics[width=\linewidth]{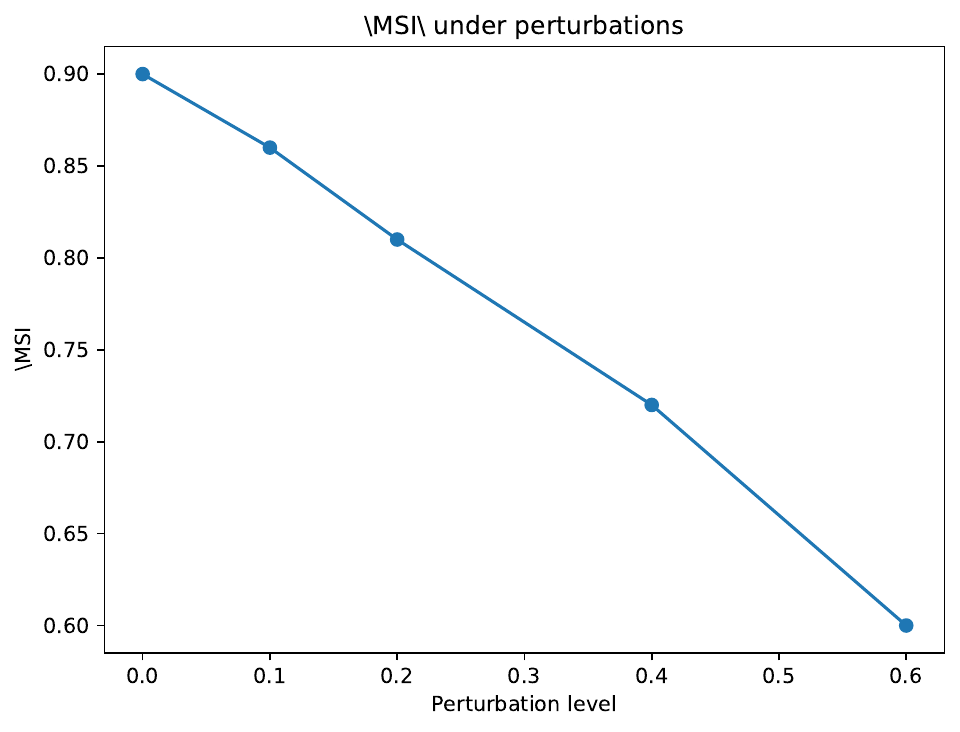}\\
    \vspace{2pt}
    \small \MSI\ under perturbations.
  \end{minipage}\hfill
  \begin{minipage}{.49\linewidth}
    \centering
    \includegraphics[width=\linewidth]{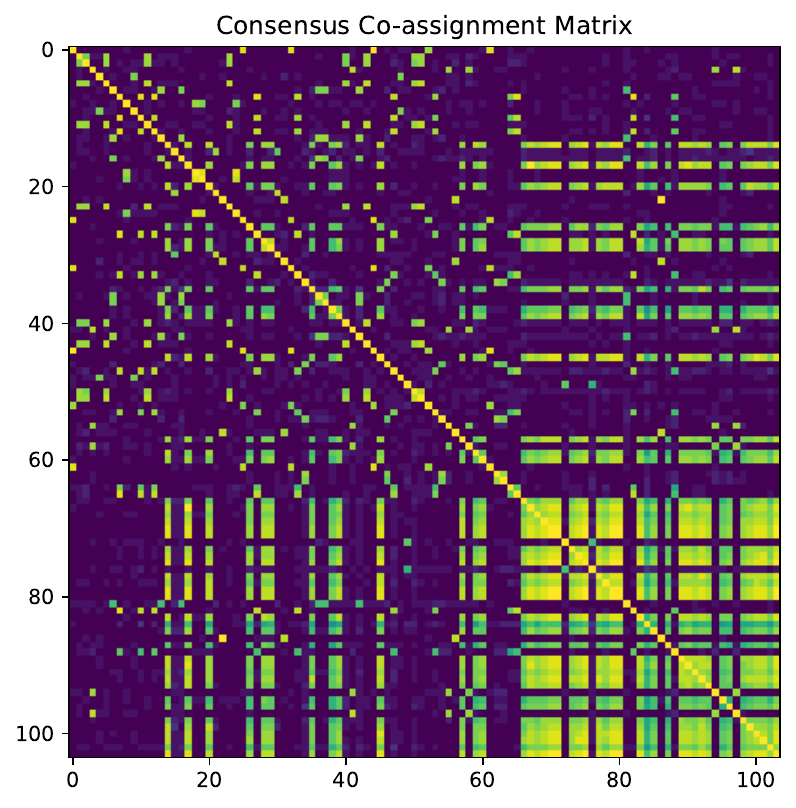}\\
    \vspace{2pt}
    \small Consensus co-assignment matrix.
  \end{minipage}
  \caption{Stability diagnostics: curves (left) and consensus (right).}
  \label{fig:viz:stability}
\end{figure}

\subsection*{Reporting templates}
\textbf{Module summary table.} Key metrics per module: size $|M|$, average degree, \RI, \BEI, mean $|\psi|$, top features, and ablation drop $\Delta_y(M)$.

\begin{table}[t]
  \centering
  \caption{Module summary. Higher \BEI{} indicates greater group-conditioned disparity; \RI{} captures redundancy.}
  \label{tab:module_summary}
  \scriptsize
  \setlength{\tabcolsep}{3pt}
  \renewcommand{\arraystretch}{1.1}
  \begin{tabular}{@{}lrrrrrr@{}}
    \toprule
    Module & $|M|$ & \shortstack{Avg\\deg.} & \shortstack{\RI\\(red.)} &
    \BEI & \shortstack{Mean\\$|\psi|$} & \shortstack{$\Delta R^2$\\(ablate $M$)} \\
    \midrule
    M0  & 11 & 0.558 & 0.214 & 0.061 & 0.014 & 0.001 \\
    M1  &  6 &10.711 & 0.372 & 1.021 & 1.247 & 0.018 \\
    M2  &  9 & 7.183 & 0.298 & 0.284 & 0.603 & 0.006 \\
    M3  &  5 &10.806 & 0.341 & 2.127 & 0.109 & 0.026 \\
    M4  &  6 & 0.959 & 0.186 & 0.117 & 0.358 & 0.002 \\
    M5  &  3 & 0.978 & 0.152 & 0.131 & 0.101 & 0.001 \\
    M6  &  7 & 1.872 & 0.205 & 0.173 & 0.029 & 0.001 \\
    M7  &  4 & 0.211 & 0.044 & 0.017 & 0.006 & 0.000 \\
    M8  &  4 & 0.667 & 0.233 & 0.471 & 0.051 & 0.004 \\
    M9  &  3 & 0.297 & 0.121 & 0.053 & 0.011 & 0.000 \\
    M10 &  3 & 0.105 & 0.037 & 0.009 & 0.003 & 0.000 \\
    M11 & 43 & 0.184 & 0.061 & 0.012 & 0.008 & 0.000 \\
    \bottomrule
  \end{tabular}
\end{table}

\subsection*{Style and reproducibility}
\textbf{Style.} Use identical color maps and legends across datasets; encode uncertainty with CIs or violin widths; prefer transparent backgrounds and vector export (PDF). 
\textbf{Reproducibility.} Each figure includes a caption noting: dataset/split, model/explainer settings, edge rule, sparsification, community algorithm, and random seed(s). We ship scripts to regenerate every figure given saved $\Phi$, $W$, and partitions.

\paragraph{Checklist (per figure).}
caption with settings \ $\bullet$ \ axis and units \ $\bullet$ \ legend keyed to modules \ $\bullet$ \ uncertainty shown \ $\bullet$ \ consistent scales \ $\bullet$ \ PDF export with embedded fonts.

\section{Limitations and Responsible Use}\label{sec:limitations}

\paragraph{Scope and assumptions.}
MoI analyzes \emph{explanation-derived} affinities between features. It is designed for tabular data with accessible per-instance attributions and aims to surface \emph{meso-scale} structure (modules). Findings depend on (i) the trained model $f$, (ii) the attribution method and its settings, and (iii) graph-construction choices. MoI is \emph{hypothesis-generating}, not a substitute for causal inference or domain oversight.

\subsection*{Methodological limitations}
\begin{enumerate}[leftmargin=*]
  \item \textbf{Attribution dependence.} Module structure varies with the explainer (e.g., SHAP vs.\ IG) and with explainer hyperparameters (background set, link function). \emph{Mitigation:} triangulate across explainers; report cross-explainer agreement and \MSI.
  \item \textbf{Background/reference sensitivity.} With SHAP/Kernel methods, changing the background $\mathcal{B}$ can shift $\Phi$ and thus $W$. \emph{Mitigation:} evaluate multiple $\mathcal{B}$ (e.g., $k$-medoids, class-/group-conditional) and include sensitivity plots.
  \item \textbf{Edge-definition bias.} Cosine emphasizes magnitude co-activation; correlations capture sign; MI/HSIC detect nonlinear ties but are noisier. \emph{Mitigation:} justify edge choice, apply shrinkage and significance filtering, and verify consistency of high-level conclusions across alternatives.
  \item \textbf{Sparsification and resolution.} Top-$k$ and thresholds control granularity; extreme settings can fragment or merge modules (resolution limit). \emph{Mitigation:} use stability-based model selection; report partitions across $(k,\theta)$ sweeps.
  \item \textbf{Stability and non-uniqueness.} Community detection is non-convex; different seeds or small data perturbations can alter partitions. \emph{Mitigation:} report \MSI, consensus matrices, and confidence intervals on module metrics.
  \item \textbf{Signed cancellations.} Summed module attributions may hide opposing signs within a module. \emph{Mitigation:} present both signed and magnitude views; visualize intra-module sign structure.
  \item \textbf{Ablation realism.} Hard “masking’’ may generate out-of-distribution inputs. \emph{Mitigation:} prefer conditional baselines (draws from $P(X_M\mid X_{\bar M})$) or soft attenuations; disclose the intervention policy.
  \item \textbf{Confounding and common causes.} Co-influence reflects associations induced by unobserved factors; modules are not inherently causal. \emph{Mitigation:} treat modules as hypotheses and follow with interventional or identification analyses when causal claims are needed.
  \item \textbf{Sample bias and shift.} Modules discovered on one cohort may not transfer. \emph{Mitigation:} evaluate across environments/time and include invariance checks; flag environment-specific modules.
  \item \textbf{Computational constraints.} Dense affinities scale as $\mathcal{O}(nd^2)$; MI/HSIC are costly. \emph{Mitigation:} pre-screen features, use sparse backbones, and report compute budgets.
\end{enumerate}

\subsection*{Fairness, privacy, and ethical use}
\begin{enumerate}[leftmargin=*]
  \item \textbf{Sensitive attributes.} Estimating \BEI\ and group-conditional effects requires careful, lawful handling of protected attributes ($A$). \emph{Practice:} apply least-privilege access, aggregate where possible, and obtain approvals where required.
  \item \textbf{Proxies and disparate treatment.} Reducing reliance on an explicit $A$ variable while retaining strong proxies in a module can \emph{increase} harm. \emph{Practice:} identify high-\BEI\ modules and address proxy pathways, not just visible attributes.
  \item \textbf{Measurement and representation harms.} Noisy or biased features (e.g., policing data) can dominate modules. \emph{Practice:} annotate data provenance and uncertainty; consider reweighting or exclusion with justification.
  \item \textbf{Privacy leakage.} Fine-grained attribution releases may reveal individual information. \emph{Practice:} publish aggregated module metrics; consider DP noise for public artifacts.
  \item \textbf{Causal claims.} Do not interpret module associations as mechanisms. \emph{Practice:} when needed, estimate path-specific or interventional effects with explicit assumptions and sensitivity analysis.
\end{enumerate}

\subsection*{Operational guidance}
\begin{itemize}[leftmargin=*]
  \item \textbf{Do} report: explainer settings, edge rule, sparsification, community algorithm, seeds, and stability diagnostics (\MSI, consensus).
  \item \textbf{Do} include: pre/post-intervention accuracy and fairness metrics with CIs; trade-off plots; ablation policies.
  \item \textbf{Don’t} deploy module-based mitigations without human review or domain sign-off.
  \item \textbf{Don’t} hide sensitive effects by dropping $A$ while keeping proxy-rich modules; disclose residual proxy risk.
\end{itemize}

\subsection*{Risk–mitigation summary}
\begin{table}[h]
  \centering
  \caption{Common risks and recommended mitigations.}
  \begin{tabular}{p{0.38\linewidth} p{0.55\linewidth}}
    \toprule
    \textbf{Risk} & \textbf{Mitigation} \\
    \midrule
    Module instability across seeds/backgrounds & Stability selection, consensus clustering, report \MSI\ and variance. \\
    Spurious associations interpreted as causal & Reserve causal language; pursue interventional/identification follow-ups. \\
    Unrealistic ablations & Use conditional baselines or soft attenuations; document policies. \\
    Proxy-induced unfairness & Rank by \BEI; intervene at module level; monitor post-intervention disparity and accuracy. \\
    Over-fragmentation/merging (resolution limit) & Hyperparameter sweeps; multi-scale analysis; select by stability and MDL/modularity targets. \\
    Privacy leakage in reports & Aggregate module statistics; suppress small cells; consider DP noise. \\
    \bottomrule
  \end{tabular}
\end{table}

\paragraph{Responsible release.}
Accompany public results with (i) a limitations note summarizing the above, (ii) reproducibility artifacts (configs, seeds, figures as PDF), and (iii) contacts for redress. MoI is most effective as part of a governance process that combines technical analysis with stakeholder input and policy oversight.

\section{Conclusion}
We introduced \emph{Modules of Influence} (MoI), a graph-based framework that elevates instance-level attributions to the meso-scale by constructing a feature–feature \emph{co-influence} graph and extracting communities as interpretable \emph{modules}. This perspective complements traditional XAI by revealing groups of features that \emph{jointly} affect predictions, enabling actions that are difficult to motivate from flat, per-feature scores alone.

\paragraph{Summary of contributions.}
MoI provides (i) a flexible recipe for building explanation graphs from diverse attribution methods, (ii) community-detection–driven modules with \emph{module-level} auditing metrics—stability (\MSI), redundancy (\RI), synergy (\Syn), and bias exposure (\BEI), and (iii) a practical evaluation protocol spanning synthetic ground truth and real tabular tasks. Visual reporting (\cref{sec:viz}) turns these analytics into decision aids via module graphs, Sankey flows, reordered heatmaps, fairness dashboards, and stability diagnostics.

\paragraph{Key findings.}
Across datasets, MoI discovers domain-aligned groups (e.g., income–education–occupation), localizes disparities to a small number of high-\BEI\ modules where targeted mitigation achieves measurable gap reductions with limited accuracy impact, and yields compact representations ($\Psi$) that preserve performance while improving parsimony. Stability analyses show that edge choices and sparsification matter; magnitude–cosine with mutual-$k$ produced robust partitions in our settings.

\paragraph{Implications.}
By shifting explanations from individual features to \emph{modules}, MoI supports concrete interventions: attenuating problematic modules, prioritizing data collection for underrepresented modules, or regularizing redundancy-heavy modules to curb overfitting. The same abstractions inform governance—audits can track a small set of module-level indicators instead of dozens of volatile feature scores.

\paragraph{Future directions.}
Promising avenues include (i) \emph{causal} follow-ups via path-specific and interventional effects at the module level, (ii) \emph{temporal} and \emph{environmental} MoI for distribution shift and monitoring, (iii) extensions beyond tabular data using concept bottlenecks or token/patch attributions, and (iv) \emph{training-time} objectives that directly encourage stable, fair module structure.

\paragraph{Closing.}
MoI encourages module-centric XAI: explanations that are robust enough to repeat, structured enough to act on, and transparent enough to audit. Treating modules as first-class citizens—rather than afterthoughts of feature rankings—opens a practical path toward trustworthy, intervention-ready model understanding.

\section*{Reproducibility Checklist (for appendix)}
\paragraph{Datasets, preprocessing, and splits.}
\begin{itemize}[leftmargin=*]
  \item \textbf{Datasets:} name, version/hash, license, download URL/date. For synthetic data, publish the generator code and fixed RNG seeds.
  \item \textbf{Preprocessing:} imputation strategy; winsorization/clipping; one-hot/ordinal encodings; standardization (per-feature mean/variance or robust alternatives); train/val/test leakage checks.
  \item \textbf{Splits:} exact indices for train/val/test (or RNG seeds $s_{\text{split}}$); stratification variables; environment/time-based splits where applicable.
\end{itemize}

\paragraph{Models and training.}
\begin{itemize}[leftmargin=*]
  \item \textbf{Architectures/params:} GBDT (trees, depth, learning rate, subsampling), RF (trees, max-features), MLP (layers, width, activation, norm, dropout).
  \item \textbf{Optimization:} optimizer, LR schedule, epochs/early stopping, batch size; class weighting; calibration method (Platt/isotonic).
  \item \textbf{Seeds/hardware:} $\text{seed}_{\text{model}}$, deterministic flags (e.g., cuDNN), device types (CPU/GPU, model), RAM/GPU RAM.
\end{itemize}

\paragraph{Attribution settings.}
\begin{itemize}[leftmargin=*]
  \item \textbf{Explainers:} SHAP (Tree/Kernel), IG (steps, baseline), LIME (kernel width, samples).
  \item \textbf{Background/reference} $\mathcal{B}$: construction (random, $k$-medoids $k=\{50,100,200\}$, class-/group-conditional), link function $\ell$ (identity/logit), output space (log-odds/probability).
  \item \textbf{Stability knobs:} number of samples for KernelSHAP/LIME; IG path discretization.
\end{itemize}

\paragraph{Graph construction.}
\begin{itemize}[leftmargin=*]
  \item \textbf{Working matrix} $A\in\{\Phi,|\Phi|\}$, column scaling (L2/MAD), optional row scaling.
  \item \textbf{Affinity rule:} cosine, (partial) correlation, MI/HSIC; implementation details (bins/kNN, kernels).
  \item \textbf{Shrinkage/significance:} shrinkage $\alpha$; permutation/FDR thresholds.
  \item \textbf{Sparsification:} $k$-NN vs.\ mutual-$k$ (report $k$), or threshold $\theta$; connectivity tweaks; symmetrization and degree normalization ($\beta$).
\end{itemize}

\paragraph{Communities and hyperparameters.}
\begin{itemize}[leftmargin=*]
  \item \textbf{Algorithms:} Louvain/Leiden (resolution, iterations), Infomap (trials), hSBM (levels/priors).
  \item \textbf{Selection:} stability-based model selection protocol (grid for $(k,\text{resolution})$, bootstrap count), objective thresholds (modularity $Q$, MDL).
\end{itemize}

\paragraph{Evaluation and metrics.}
\begin{itemize}[leftmargin=*]
  \item \textbf{Performance:} AUROC/AP or $R^2$/RMSE with 95\% CIs over seeds.
  \item \textbf{Module metrics:} \MSI\ (definition, bootstrap scheme), \RI, \Syn, \BEI; fairness metrics (DP/EO gaps) with CIs.
  \item \textbf{Ablations:} intervention policy for $do(X_M:=\tilde X_M)$ (conditional baseline generator, soft attenuation), number of samples per intervention.
\end{itemize}

\paragraph{Artifacts and scripts.}
\begin{itemize}[leftmargin=*]
  \item \textbf{Manifest:} configs (\texttt{.yaml}), logs, saved $\Phi$, $W$, partitions $\hat{\mathcal{M}}$, and figure notebooks; code commit hash.
  \item \textbf{Paths:} \texttt{scripts/compute\_phi.py}, \texttt{scripts/build\_graph.py}, \texttt{scripts/communities.py}, \texttt{scripts/ablations.py}, \texttt{viz/*.ipynb}.
  \item \textbf{Licenses:} dataset/model/third-party library licenses; usage notes.
\end{itemize}

\paragraph{Determinism \& budgets.}
\begin{itemize}[leftmargin=*]
  \item RNG seeds: $s_{\text{split}}$, $s_{\text{train}}$, $s_{\text{attr}}$, $s_{\text{graph}}$, $s_{\text{comm}}$.
  \item Compute wall-clock and peak RAM/GPU for attribution, graph, communities; environment details (OS, Python, CUDA).
\end{itemize}

\appendix
\section{Algorithms}

\begin{algorithm}[t]
\caption{Build Explanation Graph and Modules}\label{alg:build-graph}
\KwIn{$\Phi\in\R^{n\times d}$; edge rule $r$; sparsity $k$ or threshold $\theta$; community algorithm $\mathcal{C}$; options: signed/signed-layered, shrinkage $\alpha$, degree norm $\beta$}
\KwOut{Graph $G=(V,E,W)$; modules $\mathcal{M}$; module attributions $\Psi$}
\BlankLine
$A \leftarrow \Phi$ or $A\leftarrow|\Phi|$ \tcp*{choose signed or magnitude view}
Column-scale $A_{:i}\!\leftarrow\!A_{:i}/(\|A_{:i}\|_{\text{MAD or }2}+\varepsilon)$; optional row scaling \;
Compute dense affinities $\widehat{W}\gets r(A)$ \tcp*{cos/corr/pcorr/MI/HSIC}
(Optional) shrinkage: $\tilde W \leftarrow \alpha\,\widehat W + (1{-}\alpha)\,\bar w\mathbf{1}\mathbf{1}^\top$; zero small entries \;
(Optional) significance filtering via permutations (FDR control) \;
Sparsify: keep mutual-$k$ neighbors (or $|\tilde w_{ij}|\ge\theta$ with min-degree); ensure connectivity with a light $k_0$-NN backbone \;
Symmetrize: $W \leftarrow (\tilde W+\tilde W^\top)/2$; optionally degree-normalize $W\leftarrow D^{-\beta} W D^{-\beta}$ \;
Run $\mathcal{C}$ on $W$ to obtain partition $\mathcal{M}=\{M_1,\dots,M_K\}$ \;
Compute $\Psi_{sM}\leftarrow \sum_{i\in M}\phi^{(s)}_i$ for all $s,M$ \;
\Return $(G,\mathcal{M},\Psi)$\;
\end{algorithm}

\begin{algorithm}[htp]
\caption{Module Stability Index (MSI)}\label{alg:msi}
\KwIn{Graph-building config $\Gamma$; community algorithm $\mathcal{C}$; perturbation scheme $\Pi$; repetitions $T$}
\KwOut{\MSI\ and per-pair stability matrix}
\BlankLine
\For{$t\gets 1$ \KwTo $T$}{
  Sample a perturbation $\pi_t \sim \Pi$ \tcp*{bootstrap rows, vary background $\mathcal{B}$, noise on $A$}
  Build $W^{(t)}$ with config $\Gamma$ under $\pi_t$ \;
  Compute partition $\mathcal{M}^{(t)}$ using $\mathcal{C}$ \;
}
Compute consensus matrix $C_{ij} \leftarrow \frac{1}{T}\sum_t \mathbb{1}[c_i^{(t)}=c_j^{(t)}]$ \;
Match modules across runs with Hungarian assignment on $1{-}\mathrm{IoU}$ between sets \;
\MSI\ $\leftarrow$ mean IoU of matched module pairs (report mean $\pm$ CI) \;
\Return \MSI, $C$ \;
\end{algorithm}

\begin{algorithm}[htbp]
\caption{Module Ablation and Synergy}\label{alg:ablation}
\KwIn{Trained predictor $f$; data $x^{(s)}$; modules $\mathcal{M}$; attribution matrix $\Phi$; intervention policy $\pi$ (hard or soft); evaluation metric $\mathcal{E}$}
\KwOut{Ablation drops $\Delta_y(M)$; synergy scores $\Syn(A,B)$}
\BlankLine
\ForEach{module $M\in\mathcal{M}$}{
  \ForEach{instance $x^{(s)}$}{
    Construct counterfactual $x^{(s,M)} \sim \text{do}_\pi(X_M:=\tilde X_M \mid X_{\bar M}{=}x^{(s)}_{\bar M})$ \tcp*{conditional baseline or attenuation}
    $\hat y^{(s)} \leftarrow f(x^{(s)})$, \quad $\hat y^{(s,M)} \leftarrow f(x^{(s,M)})$ \;
  }
  $\Delta_y(M)\leftarrow \mathcal{E}(\{\hat y^{(s)}\}) - \mathcal{E}(\{\hat y^{(s,M)}\})$ \;
}
\ForEach{pair $(A,B)$}{
  Similarly compute $\Delta_y(A\cup B)$ using joint intervention \;
  $\Syn(A,B) \leftarrow \Delta_y(A\cup B) - \Delta_y(A) - \Delta_y(B)$ \tcp*{Eq.~\ref{eq:syn}}
}
\Return $\{\Delta_y(M)\}, \{\Syn(A,B)\}$ \;
\end{algorithm}

\section{Implementation Notes}\label{sec:impl}
We release a Python package that exposes pluggable \emph{edge rules}, \emph{sparsifiers}, and \emph{community algorithms} (Louvain/Leiden/Infomap/hSBM), plus utilities for \BEI, \RI, \MSI, and \Syn, and a visualization layer (graphs/heatmaps/Sankey/fairness dashboard).

\subsection*{Package layout}
\begin{itemize}[leftmargin=*]
  \item \texttt{moi/graphs.py} — edge rules, shrinkage, significance, sparsifiers, symmetrization, degree normalization.
  \item \texttt{moi/community.py} — wrappers for Louvain/Leiden/Infomap/hSBM; stability selection utilities.
  \item \texttt{moi/metrics.py} — \RI, \BEI, \MSI, \Syn, modularity $Q$, conductance, ARI/NMI, VI.
  \item \texttt{moi/ablations.py} — hard/soft module interventions; conditional baselines.
  \item \texttt{moi/attr.py} — attribution IO helpers (SHAP/IG/LIME outputs $\rightarrow \Phi$), background construction.
  \item \texttt{moi/viz.py} — module graphs, reordered heatmaps, Sankey, fairness dashboard, stability plots.
  \item \texttt{moi/io.py} — read/write $\Phi$, $W$, partitions $\hat{\mathcal{M}}$; GraphML/CSV/NPZ; figure exporters (PDF).
  \item \texttt{cli/} — command-line entry points (see below).
  \item \texttt{examples/} — end-to-end notebooks (synthetic, tabular fairness).
\end{itemize}

\subsection*{Core API}
\noindent Minimal fit/transform interface:
\begin{lstlisting}[style=ieeecode]
from moi import MoI

moi = MoI(
    edge_rule="cosine_mag",        # cosine|corr|pcorr|mi|hsic
    k=20, mutual=True,             # sparsification
    signed=False, degree_norm=0.5, # graph normalization
    community="leiden",            # louvain|leiden|infomap|hsbm
    resolution=1.0, random_state=0
)

modules, Psi, graph = moi.fit(Phi)  # Phi: (n,d) attributions
scores = moi.metrics()              # RI, BEI, MSI, Syn, Q, ...
moi.save("artifacts/run01/")
\end{lstlisting}

\noindent Module-level ablation:
\begin{lstlisting}[style=ieeecode]
from moi.ablations import ablate_modules
drops, synergy = ablate_modules(
    model=f, X=X_test, modules=modules,
    policy="conditional", generator=cond_model
)
\end{lstlisting}

\subsection*{Edge rules \& sparsifiers}
\textbf{Edge rules:} \texttt{cosine\_mag}, \texttt{corr\_signed}, \texttt{pcorr\_signed}, \texttt{mi\_knn}, \texttt{hsic\_rbf}. \\
\textbf{Shrinkage/significance:} $\tilde W \!\leftarrow\! \alpha\,\widehat W{+}(1{-}\alpha)\bar w$, permutation FDR. \\
\textbf{Sparsifiers:} \texttt{topk}, \texttt{mutual\_topk}, \texttt{threshold}, optional $k_0$-NN backbone; symmetrize and degree-normalize ($\beta \!\in\! \{1/2,1\}$).

\subsection*{Communities \& stability}
Wrappers expose common knobs (\texttt{resolution}, \texttt{trials}, \texttt{levels}). \MSI\ is computed via bootstrap resamples with Hungarian matching on IoU; consensus matrices are optionally returned. A stability-driven selector sweeps $(k,\text{resolution})$ and picks Pareto-optimal settings (maximize \MSI\ subject to $Q$/MDL thresholds).

\subsection*{Metrics}
\begin{itemize}[leftmargin=*]
  \item \textbf{\RI} (redundancy): mean $|\mathrm{corr}|$ within modules on $A\!\in\!\{\Phi,|\Phi|\}$.
  \item \textbf{\BEI} (bias exposure): group-conditional $\psi^{(s)}(M)$ contrasts with pooled-variance denominator; CIs via bootstrap.
  \item \textbf{\MSI} (stability): mean IoU of matched modules across perturbations; report mean$\pm$CI.
  \item \textbf{\Syn} (synergy): super-additivity under joint interventions; pairwise table and optional higher-order scans.
\end{itemize}

\subsection*{Visualization}
\texttt{viz.module\_graph(W, modules, signed=True, pdf=True)}, \\
\texttt{viz.heatmap\_W(W, modules)}, \\
\texttt{viz.sankey\_flows(Phi, modules, by="class|group")}, \\
\texttt{viz.fairness\_dashboard(BEI, disparities, deltas)}, \\
\texttt{viz.stability\_curves(msi\_by\_perturb)}. \\
All figures export as vector PDF with consistent color maps; negative edges/attributions shown with diverging palettes or dashed overlays.

\subsection*{CLI}
\begin{verbatim}
moi build-graph   --phi phi.npz --edge cosine_mag --k 20 --mutual \
                  --degree-norm 0.5 --out artifacts/run01/
moi communities   --graph artifacts/run01/W.npz --algo leiden --res 1.0
moi metrics       --phi phi.npz --modules modules.json --labels groups.csv
moi ablate        --model model.pkl --X X_test.npz --modules modules.json \
                  --policy conditional --out artifacts/run01/ablations/
moi visualize     --graph ... --modules ... --out figs/
\end{verbatim}

\subsection*{Performance \& scalability}
\begin{itemize}[leftmargin=*]
  \item \textbf{ANN/backbone:} approximate top-$k$ neighbors for cosine/corr; fall back to exact for small $d$.
  \item \textbf{Sparse ops:} store $W$ as CSR; most community backends accept sparse matrices.
  \item \textbf{Batching:} compute $\Phi$ and MI/HSIC in batches; pre-screen pairs by variance/dot-product thresholds.
  \item \textbf{Complexity:} cosine/corr $\tilde{\mathcal{O}}(ndk)$ with ANN; memory $\Theta(dk)$ after sparsification.
\end{itemize}

\subsection*{Reproducibility}
Deterministic seeds (\texttt{split/train/attr/graph/comm}); YAML configs saved with every artifact; logs include OS/Python/BLAS/GPU details, wall-clock, and peak RAM.

\subsection*{File formats}
\texttt{phi.npz} (CSR or dense, shape $n{\times}d$), \texttt{W.npz} (sparse), \texttt{modules.json} (list of index arrays), \texttt{consensus.npz}, GraphML export (\texttt{graph.graphml}), and PDF figures.

\subsection*{Extensibility}
\textbf{New edge rules:} implement \texttt{EdgeRule.fit(A)\,$\to\,W$}. \\
\textbf{New community methods:} subclass \texttt{Community.fit(W)\,$\to\,\mathcal{M}$}. \\
\textbf{Custom fairness metrics:} register functions with signature \texttt{f(y, \={y}, A)\,$\to$\,scalar} for dashboards.

\subsection*{Example config}
\begin{verbatim}
edge_rule: cosine_mag
signed: false
column_scaling: MAD
sparsifier: mutual_topk
k: 20
degree_norm: 0.5
community: leiden
resolution: 1.0
stability:
  bootstraps: 200
  res_sweep: [0.5, 1.0, 1.5]
  k_sweep: [10, 20, 30]
fairness:
  group_label: A
  bei_eps: 1e-6
\end{verbatim}
```





\vfill


\bibliographystyle{IEEEtran}
\bibliography{IEEEabrv}

\end{document}